\documentclass[a4paper]{article}

\usepackage{amsmath,amssymb,amsfonts} 
\usepackage[mathcal]{eucal}
\usepackage{layout} 

\newtheorem{Lemma}{Lemma}

\newtheorem{Corollary}{Corollary}
\newtheorem{Proposition}{Proposition}

\date{}
\title{Subdiffusive behavior generated by irrational rotations}
\author{F. Huveneers\footnote{Partially supported by the Belgian IAP program P6/02 and by the University of Helsinki.}\\
UcL, FYMA, 2 Chemin du Cyclotron, \\
B-1348 Louvain-la-Neuve, Belgium.\\
E-mail : francois.huveneers@uclouvain.be}

\begin{document}

\maketitle

\begin{abstract}
We study asymptotic distributions of the sums $y_n(x)=\sum_{k=0}^{n-1}\psi (x+k\alpha)$ with respect to the Lebesgue measure, 
where $\alpha\in\mathbf R-\mathbf Q$
and where $\psi$ is the 1-periodic function of bounded variation such that $\psi (x)=1$ if $x\in\lbrack 0,1/2\lbrack$ and $\psi(x)=-1$ if $x\in\lbrack 1/2,1\lbrack$.
For every $\alpha\in\mathbf R-\mathbf Q$, we find a sequence $(n_j)_j\subset \mathbf N$ such that $y_{n_j}/\sqrt j$ is asymptotically normally distributed.
For $n\ge 1$, let $z_n\in(y_m)_{m\le n }$ be such that $||z_n||_{\mathrm L^2}=\max_{m\le n }||y_m||_{\mathrm L^2}$. 
If $\alpha$ is of constant type, we show that $z_n/||z_n||_{\mathrm L^2 }$ is also asymptotically normally distributed.
We give an heuristic link with the theory of expanding maps of the interval. 
\end{abstract}

\section{Introduction}

Some purely deterministic, smooth and finite dimensional dynamical systems may generate diffusion process.
Such a diffusion is due to uncertainty on initial conditions.
If a distribution is initially concentrated in one point, it will remain so under the flow of such a system. 
But if the initial conditions are distributed on some larger set of the phase space, it may well be that the distribution evolves diffusively.

Some cases of deterministic diffusion have been successfully investigated \cite{bal}. 
Let us mention the theory of expanding maps of the interval \cite{hof}, and the important result by Bunimovich and Sinai about the Lorentz gas \cite{bun}.
In the two previous examples, the underlying dynamical system is hyperbolic ;  and
it has been suggested that macroscopic diffusion is generally due to microscopic chaos \cite{gas}. 
But numerical experiments with systems of zero Lyapunov exponents show that 
diffusion may happen even in the absence of hyperbolicity \cite{det}. 

The rotation of the circle by an irrational angle is a well known example of ergodic non hyperbolic dynamical system.
Burton and Denker \cite{bur} (see also \cite{del})
have shown that one may find a function $\psi\in\mathrm L^2(\mathbf T,\mathbf R)$ such that $y_n/||y_n||_{\mathrm L^2}$ is asymptotically normally distributed.
By Denjoy-Koksma inequality (see (\ref{DenjoyKoksma}) below), this $\psi$ is not a bounded variation function.
Among other results, 
Liardet and Voln\'y
have shown  (Theorems 1 and 2 in \cite{lia}) that, if $r\ge 0$, 
then there exist numbers $\alpha\in\mathbf R-\mathbf Q$ and 
a sequence $(d_n)_n\subset\mathbf R_0^+$ such that
for every $\psi$ in a dense $\mathrm G_\delta$ set of $\mathcal C^r(\mathbf T,\mathbf R)$, 
the distributions of $d_n y_n$ form a dense set in the space of all probability measures on $\mathbf R$. 
Their results do not cover the case where $\alpha$ is of constant type (see (\ref{constanttype}) below) and $\psi$ of bounded variation.

Let $\mathbf T=\mathbf R/\mathbf Z$. If $u\in\mathrm L^1(\mathbf T,\mathbf R)$, one defines 
\begin{equation}\label{Var}
\mathrm{Var}(u)=\sup\lbrace\int_0^1uv'\mathrm d x : v\in\mathcal C^1(\mathbf T,\mathbf R), ||v||_{\mathrm L^\infty}\le 1\rbrace.
\end{equation}
One defines also the set $\mathrm{BV}(\mathbf T,\mathbf R)=\lbrace u\in \mathrm L^1(\mathbf T,\mathbf R) :\mathrm{Var}(u)<\infty\rbrace$.

Let $\psi\in\mathrm{BV}(\mathbf T,\mathbf R)$ be such that $\int_0^1\psi\mathrm dx=0$. Let $\alpha\in\mathbf R - \mathbf Q$. We consider the map
\begin{equation}\label{system}
\mathrm F : \mathbf T\times\mathbf R \rightarrow \mathbf T\times\mathbf R : (x,y)\mapsto (x+\alpha,y+\psi (x)).
\end{equation}
If $n\in\mathbf N$, 
one defines implicitly the function $y_n\in\mathrm{BV}(\mathbf T,\mathbf R)$ by the relation
\begin{equation}\label{yn}
\mathrm F^n(x,y)=(x+n\alpha,y+y_n(x)) .
\end{equation}
Explicitly, one has
$y_n(x)=\sum_{k=0}^{n-1}\psi(x+k\alpha)$ for $n\ge 1$. Although $y_n$ depends on $\psi$ and $\alpha$, one will not generally write it. 
Let $m_\mathrm L$ be the Lebesgue measure on $\mathbf T$.
The space $(\mathbf T,m_\mathrm L)$ is then a probability space, and  $(y_n)_{n\ge 0}$ is a sequence of random variables on this space. 

The sequence $(y_n)_{n\ge 0}$ has been widely studied \cite{ada}\cite{drm}\cite{her}\cite{kui}. Here are two important informations. 
First, the sequence $(y_n)_{n\ge 0}$ is bounded in $\mathrm L^2(\mathbf T,\mathbf R)$ if and only if there exists $u\in\mathrm L^2(\mathbf T,\mathbf R)$ such that $\mathrm R_\alpha u - u =\psi$ (where  by definition $\mathrm R_\alpha u(x)=u(x+\alpha)$) (\cite{her} p.183). 
Next,  let $p/q$ be an irreducible fraction such that $|\alpha -p/q|\le 1/q^2$ (by Dirichlet theorem, there are infinitely many such fractions). 
Denjoy-Koksma inequality (\cite{her} p.73) asserts that 
\begin{equation}\label{DenjoyKoksma}
||y_q||_{\mathrm L^\infty} = ||\sum_{k=0}^{q-1}\mathrm R_{k\alpha }\psi ||_{\mathrm L^\infty }\le\mathrm{Var}(\psi).
\end{equation}

Let us now present our results. 
We will actually only consider the function $\psi_*$ defined by 
\begin{equation}\label{psistar}
\psi_*(x)=1\quad \mathrm{if}\quad 0\le x<1/2, \qquad  \psi_*(x)=-1 \quad\mathrm{if}\quad 1/2\le x<1. 
\end{equation}
It is known that there is no $u\in\mathrm L^2(\mathbf T,\mathbf R)$ that solves the equation $\mathrm R_\alpha u -u = \psi_*$ 
(Lemma \ref{cohomologic}, Section \ref{sec2}).

First, can we find an increasing sequence $(n_j)_{\ge 1}\subset\mathbf N$ such that $y_{n_j}/\sqrt j$ should be asymptotically normally distributed (with strictly positive variance) ? 
Proposition \ref{result1} answers this question positively. 
This means that, if we looked at the system at the times $n_j$ only, we should observe a diffusion process.
Next, how fast has to grow the sequence $(n_j)_{j\ge 1}$ ? 
If $\alpha$ is of constant type (see (\ref{constanttype})), we will see  that it may be taken to grow exponentially, but not slower
(see Remark after Proposition \ref{result3}, and Corollary \ref{result2}). 

It seems also natural to consider the sequence $(z_n)_{n\ge 0 }\subset\mathrm{BV}(\mathbf T,\mathbf R)$, defined as follows : 
\begin{equation}\label{zj}
z_n\in(y_m)_{0\le m\le n} \quad : \quad ||z_n||_{\mathrm L^2 } = \max_{0\le m \le n }||y_m||_{\mathrm L^2 }  
\end{equation}
(one take the first element of $(y_m)_{0\le m\le n }$ if there is more than one possibility).
In Proposition \ref{result3}, we will see that the sequence $z_n/||z_n||_{\mathrm L^2 }$ is asymptotically normally distributed.

Let $\mathrm G(\sigma)$ be the probability measure on $\mathbf R$ that admits the density $f(x)=e^{-x^2/2\sigma^2}/\sqrt{2\pi\sigma^2}$ ($\sigma >0$).
\begin{Proposition}\label{result1}
Let $(y_n)_{n\ge 0 }$ be defined in (\ref{yn}).
If $\psi = \psi_*$ (see (\ref{psistar})), and if $\alpha\in\mathbf R - \mathbf Q$,
there exists an increasing sequence $(n_j)_{j\ge 1}\subset{\mathbf N}$ such that $y_{n_j}/\sqrt{j}\stackrel{\mathrm D}{\rightarrow}\mathrm G(1)$ as $j\rightarrow\infty$.
\end{Proposition}
This result is quite weak, because the sequence $(n_j)_{j\ge 1}$ is completely unknown. Nevertheless, we believe it has some interest. First, the result is valid for any irrational number $\alpha$. Next, the proof is not technical but contains the principal ideas we need for proving our second Proposition. 
Finally, 
it allows us to make an heuristic link between our case and the theory of expanding maps of the interval
(see Section \ref{sec2}, after the proof of Proposition \ref{result1}).

One will then need the theory of continued fractions. 
Let $(a_n)_{n\ge 0}\subset\mathbf N$ be the sequence of partial quotients of $\alpha$ (see for example \cite{har} for definition and details).
The sequence $(p_n/q_n)_{n\ge 0}\subset\mathbf Q$ of convergents of $\alpha$ is then defined as follows : $p_0/q_0 = a_0/1$, $p_1/q_1 = (a_0a_1 + 1)/a_1$, and, for $n\ge 1$, 
\begin{equation}\label{convergents}
q_{n+1} = a_{n+1}q_n + q_{n-1}, \quad 
p_{n+1} = a_{n+1}p_n + p_{n-1}.
\end{equation}
One will usually not write explicitly the dependence of $a_n$ and $p_n/q_n$ on $\alpha$.
Here is a fundamental result of the theory of continued fractions : for $n\ge 0$, one has
\begin{equation}\label{fundamentalFracCont}
\frac{1}{q_n + q_{n+1}}\le |q_n\alpha - p_n |\le \frac{1}{q_{n+1}}\le \frac{1}{a_{n+1}q_n}.
\end{equation}

Let us introduce a particular class of numbers. One says that $\alpha\in\mathbf R-\mathbf Q$ is of \emph{constant type} if 
\begin{equation}\label{constanttype}
\exists \mathrm C>0 : \forall q\in\mathbf Z_0, \forall p\in\mathbf Z , |q\alpha - p|\ge\frac{\mathrm C}{|q|}.
\end{equation}
Equivalently, $\alpha$ is of constant type if 
\begin{equation}\label{ead}
\exists d\ge 1 : \forall n\ge 0, a_n\le d.
\end{equation}
This implies that the sequence $(q_n)_{n\ge 0}$ grows only exponentially with $n$. These numbers form a set of zero Lebesgue measure. 

\begin{Proposition}\label{result3}
Let $(z_n)_{n\ge 0}$ be defined in (\ref{zj}).
Let $\psi = \psi^*$ be defined in (\ref{psistar}). 
Let $\alpha$ be a number of constant type. 
One has $z_n/||z_n||_{\mathrm L^2 }\stackrel{\mathrm D }{\rightarrow }\mathrm G(1)$ as $n\rightarrow\infty$.
Moreover, there exist $\mathrm C, \epsilon>0$ such that, if $q_j \le n < q_{j+1}$, one has $\epsilon \sqrt j \le ||z_n||_{\mathrm L^2}\le \mathrm C\sqrt j$.
\end{Proposition}
\emph{Remark.}
Let $\psi = \psi^*$, 
and let $\alpha$ be a number of constant type. 
Let $n_j$ be such that $z_{q_j}=y_{n_j}$. 
One has $n_j\le q_j$.
Let $\sigma_j = ||y_{n_j}||_{\mathrm L^2 }/\sqrt j$.
By Proposition \ref{result3}, one has $\epsilon \le \sigma_j = ||z_{q_j}||_{\mathrm L^2 }/\sqrt j\le \mathrm C$, 
and $y_{n_j}/\sqrt j \sigma_j = z_{q_j}/||z_{q_j}||_{\mathrm L^2}\stackrel{\mathrm D }{\rightarrow }\mathrm G(1)$.
\begin{Corollary}\label{result2}
Let $(y_n)_{n\ge 0 }$ be defined in (\ref{yn}).
Let $\psi = \psi_*$ be defined in (\ref{psistar}).
Let $\alpha$ be a number of constant type.
Let $(n_j)_{j\ge 1 }\subset \mathbf N$ and let $(\sigma_j)_{j\ge 1 }\subset\mathbf R^+_0$ be such that 
$y_{n_j}/\sigma_j\sqrt j\stackrel{\mathrm D}{\rightarrow }\mathrm G(1)$.
Moreover, suppose that there exist $\mathrm C>\epsilon >0$ such that $\epsilon \le \sigma_j \le \mathrm C$
for every $j\ge 1$.
Then, the sequence $(n_j)_{j\ge 1 }$ does not grow slower than exponentially with $j$.
\end{Corollary}

\noindent\emph{Question.}
What happens when $\psi\ne\psi_*$ ? 
The choice $\psi = \psi_*$ is only needed to prove $||z_n||_{\mathrm L^2 }\ge \epsilon\sqrt j$ when $n\ge q_j$. 
(Lemma \ref{minore}, Section \ref{sec4}). 
It follows from the proof of this Lemma that other choices should be possible.

The organization of the paper is as follows. Proposition \ref{result1} is shown in Section 2. In Section 3, one shows an abstract central limit theorem ; this Section is independent of the others. 
One proves Proposition \ref{result3} and Corollary \ref{result2} in Section 4. 

The letter $\mathrm C$ is used to denote a strictly positive constant that may vary from place to place.

\section{Proof of Proposition \ref{result1}}\label{sec2}

Let $\alpha\in\mathbf R - \mathbf Q$.
Let $(p_n/q_n)_{n\ge 0}$ be its convergents, and $(a_n)_{n\ge 0}$ its partial quotients. 
Let $\psi\in\mathrm{BV}(\mathbf T,\mathbf R)$ be such that $\int_0^1\psi\mathrm d x = 0$.

\begin{Lemma}\label{fraccont}
Let $n\ge 0$.

$1)$ Of the fractions $p_n/q_n$ et $p_{n+1}/q_{n+1}$, one at least satisfies $|\alpha-p/q|<1/2q^2$.

$2)$ If $q_n$ is even, then $q_{n+1}$ is odd.

$3)$ If $q_n$ and $q_{n+2}$ are even, then $|\alpha - p_{n+1}/q_{n+1}|<1/2q_{n+1}^2$.

$4)$ From four consecutive convergents, one at least has an odd denominator and satisfies the inequality 
$|\alpha - p/q|<1/2q^2$.
\end{Lemma} 
\emph{Proof}. 
For 1), see \cite{har} p.152. 
Let us show 2) by contradiction.
Let us suppose we have found a smallest 
 $j\in\mathbf N$ such that $q_j$ and $q_{j+1}$ are even. We have $j\ge 1$ and therefore $q_{j+1}=a_{j+1}q_j+q_{j-1}$. Because $q_{j-1}$ is odd and $q_j$ is even, $q_{j+1}$ should also be odd. 
Let us show 3). By 2), $q_{n+1}$ is odd, and on the other hand we have that $q_{n+2}=a_{n+2}q_{n+1}+q_n$. The number $a_{n+2}$ has to be even, and therefore $a_{n+2}\ge 2$. 
The result follows from (\ref{fundamentalFracCont}). 
Finally, 4) is obtained by considering all the possibilities. $\square$

If $u\in\mathrm L ^2(\mathbf T,\mathbf R)$, if $k\in\mathbf Z$, one writes $\hat u(k)=\int_0^1 u(x)e^{-2i\pi kx}\mathrm d x$.
If $u\in\mathrm{BV}(\mathbf T,\mathbf R)$, it follows from (\ref{Var}) that $|\hat u(k)|\le \mathrm{Var}(u)/2\pi |k|$ for $k\ne 0$.
One has 
\begin{equation}\label{Fourier}
\widehat{y_n}(k) = \frac{1-e^{2i\pi n k \alpha}}{1-e^{2i\pi k\alpha}}\hat\psi (k), \quad (n\ge 1,k\in\mathbf Z_0).
\end{equation}
Let us also introduce the following notation : if $x\in\mathbf R$, one writes 
\begin{equation}\label{distance}
|x|_\mathbf T = \inf_{p\in\mathbf Z}|x-p|.
\end{equation}
One checks the two following inequalities : for all $x,y\in\mathbf R$, one has
\begin{equation}\label{normequ}
4|x|_\mathbf T\le |1-e^{2i\pi x}|\le2\pi |x|_\mathbf T, 
\end{equation}
\begin{equation}\label{triangle}
|x+y|_{\mathbf T } \le |x|_{\mathbf T } + |y|_{\mathbf T } \quad\mathrm{and}\quad |1 - e^{2i\pi (x+y) }| \le |1- e^{2i\pi x }| + |1 - e^{2i\pi y }|.
\end{equation}
Therefore, for every $m\in\mathbf Z$,
\begin{equation}\label{linearequ}
|1-e^{2i\pi mx}|\le |m|.|1-e^{2i\pi x}|.
\end{equation}
Moreover, if $n\ge 1$, $|q_n\alpha - p_n |=|q_n\alpha |_\mathbf T$ (\cite{her} p.63). 
 
\begin{Lemma}\label{cohomologic}
Let $\psi = \psi_*$ given by $(\ref{psistar})$. There exists no $u\in\mathrm L^2(\mathbf T,\mathbf R)$ such that $\mathrm R_\alpha u - u = \psi_*$.
\end{Lemma}
\emph{Proof.}
A solution $u$ should be such that $\hat u(k) = \hat\psi_*(k)/(e^{2i\pi k\alpha}-1)=-2i/\pi k(e^{2i\pi k\alpha}-1)$ if $k$ is odd.
By point 2) of Lemma \ref{fraccont}, 
for infinitely many odd $k$, one may write $k=q_j$ for some $j\ge 1$. 
But one has $|k\alpha|_\mathbf T\le 1/|k|$ for those $k$. 
Therefore $\hat u(k)$ should not go to 0 as $k\rightarrow\infty$. $\square$

\begin{Lemma}\label{faible}
One has $y_{q_n}\rightharpoonup 0$ in $\mathrm L^2(\mathbf T,\mathbf R)$ as $n\rightarrow\infty$. 
\end{Lemma}
\emph{Proof}.
By Denjoy-Koksma inequality (\ref{DenjoyKoksma}), $||y_{q_n}||_{\mathrm L^2} \le  ||y_{q_n}||_{\mathrm L^\infty}\le\mathrm{Var}(\psi)$. Therefore, we only need to check that, if $k\in\mathbf Z_0$,
$\widehat{y_{q_n}}(k)\rightarrow 0$ as $n\rightarrow\infty$. By (\ref{Fourier}), 
\[
|\widehat{y_{q_n}}(k)|\le\frac{\mathrm{Var}(\psi)}{2\pi |k|}\frac{1}{|1- e^{2i\pi k\alpha }|} |1-e^{2i\pi q_n k\alpha}|
\]
if $k\ne 0$.
But by (\ref{linearequ})
\[
|1-e^{2i\pi q_n k\alpha}|\le |k|.|1-e^{2i\pi q_n\alpha}|\le 2\pi |k|.|q_n\alpha|_\mathbf T\rightarrow 0 
\quad \mathrm{as}\quad n\rightarrow\infty. \quad\square
\]
A direct consequence of this Lemma is that, for every $\beta\in\mathbf R$, $\mathrm R_\beta y_{q_n}\rightharpoonup 0$ in $\mathrm L^2(\mathbf T,\mathbf R)$ as $n\rightarrow\infty$.

If $x\in\mathbf R$, one sets $\overline x = x - \lfloor x\rfloor$.
Following \cite{her} p.64, 
we give some informations about some finite sequences  $(\overline{n\alpha})_n$. 
If $p/q\in\mathbf Q$ is irreducible, one has $\lbrace\overline{j.p/q} \rbrace_{0\le j\le q-1 } = \lbrace j/q \rbrace_{0\le j\le q-1 }$.
We say that $p/q\in\mathbf Q$ ($p/q$ irreducible) is a \emph{rational approximation} of $\alpha$ for the constant $0<\beta\le 1$ if the inequality $|\alpha -p/q |<\beta/q^2$ is satisfied.
Let us write $\lbrace \overline{j\alpha}\rbrace_{0\le j\le q-1 } = \lbrace \alpha_j\rbrace_{0\le j\le q-1 }$, where $0=\alpha_0<\alpha_1 <\dots<\alpha_{q-1}<1$. 

If $\alpha>p/q$, one has $k\alpha -k.p/q<k\beta/q^2<1/q$ if $1\le k\le q-1$. 
Therefore, if $0\le j \le q-1$, there exists some $l(j)\in\mathbf N$ such that $0\le \alpha_j - l(j)/q\le 1/q$. 
But the sequence  $\lbrace \alpha_j\rbrace_{0\le j\le q-1 }$ is ordered, and so $l(j)=j$. One may thus write
\begin{equation}\label{longinequ1}
0 = \alpha_0 < \frac{1}{q}<\alpha_1 < \frac{2}{q}<\alpha_2 <\dots < \frac{q-1}{q}<\alpha_{q-1} < 1. 
\end{equation}
Similarly, if $\alpha < p/q$, one has
\begin{equation}\label{longinequ2}
\alpha_0 = 0 < \alpha_1 < \frac{1}{q} < \alpha_2<\frac{2}{q} <\dots < \alpha_{q-1}<\frac{q-1}{q}<1. 
\end{equation}
In both cases one has
\begin{equation}\label{inequbeta}
|\alpha_j - j/q |< \beta/q \quad (1\le j\le q-1).
\end{equation}
The following Lemma gives a slight improvement of Denjoy-Koksma (\ref{DenjoyKoksma}) inequality when $\psi = \psi_*$ (\ref{psistar}).
\begin{Lemma}\label{Denjoy}
Let $\psi = \psi_*$ given by $(\ref{psistar})$. 
Let $p/q$ be a rational approximation of $\alpha$ for the constant $\beta\le 1/2$, 
and suppose that $q$ is odd.
Then the function $y_q$ takes only the values $\pm 1$.
\end{Lemma} 
\emph{Proof}. 
Let $\phi = \sum_{k=0}^{q-1}\mathrm R_{kp/q}\psi =\sum_{k=0}^{q-1}\mathrm R_{k/q}\psi$.
One has $\phi(x)=\psi (qx)$ ; 
indeed, 
one has $\mathrm R_{1/q}\phi =\phi$ and $\phi |_{\lbrack 0,1/q\lbrack} = (q-1)/2 +(\mathrm R_{(q-1)/2}\psi )|_{\lbrack 0,1/q\lbrack} - (q-1)/2$. 
Let us then write $\lbrace \overline{j\alpha}\rbrace_{0\le j\le q-1 } = \lbrace \alpha_j\rbrace_{0\le j\le q-1 }$, where $0=\alpha_0<\alpha_1 <\dots<\alpha_{q-1}<1$. 
One has $y_q=\sum_{k=0}^{q-1}\mathrm R_{k\alpha }\psi=\sum_{k=1}^{q-1}(\mathrm R_{\alpha_k} - \mathrm R_{k/q})\psi + \phi$.  

By (\ref{longinequ1}) and if $\alpha>p/q$, one has, for $0\le k \le q-1$,
\[
(\mathrm R_{\alpha_k} - \mathrm R_{k/q})\psi (x) = 
\left\lbrace\begin{array}{ccl}
+2 &\mathrm{if} & x\in\lbrack 1- \alpha_k , 1-k/q\lbrack,\\
-2 &\mathrm{if} & x\in\lbrack 1/2 - \alpha_k, 1/2 - k/q \lbrack   \quad (\mathrm{mod} 1),\\
0 & \mathrm{otherwise.}&\\
\end{array}\right.
\]
Similarly, if $\alpha <p/q$, one has by (\ref{longinequ2}) that, for $0\le k\le q-1$,
\[
(\mathrm R_{\alpha_k} - \mathrm R_{k/q})\psi (x) = 
\left\lbrace\begin{array}{ccl}
-2 &\mathrm{if} & x\in\lbrack 1- k/q , 1-\alpha_k\lbrack,\\
+2 &\mathrm{if} & x\in\lbrack 1/2 - k/q, 1/2 - \alpha_k \lbrack   \quad (\mathrm{mod} 1),\\
0 & \mathrm{otherwise.}&\\
\end{array}\right.
\]

One now computes $y_q$. To fix the ideas, let us consider the case $\alpha >p/q$. 
If  $0\le j \le q-1$, one has
\begin{eqnarray*}
y_q\big|_{\lbrack j/q,(j+1)/q\lbrack} 
&=&
\sum_{k=1}^{q-1}(\mathrm R_{\alpha_k} - \mathrm R_{k/q}) \psi\big|_{\lbrack j/q,(j+1)/q\lbrack} + \phi\big|_{\lbrack j/q,(j+1)/q\lbrack} \\
&=&
(2\chi_{\lbrack \frac{j+1}{q} -\delta_1(j), \frac{j+1}{q} \lbrack } -2\chi_{\lbrack \frac{j}{q} + \frac{1}{2q} -\delta_2(j) , \frac{j}{q} + \frac{1}{2q}\lbrack })
+ (\chi_{\lbrack\frac{ j}{q},\frac{j}{q} + \frac{1}{2q}\lbrack } - \chi_{\lbrack \frac{j}{q} + \frac{1}{2q}, \frac{j+1}{q}\lbrack }),
\end{eqnarray*}
where, by (\ref{inequbeta}), $0\le \delta_1(j),\delta_2(j) < 1/2q$.  $\square$

One checks that, if $u : \mathbf T\rightarrow\mathbf R$, 
if $n\ge 1$, 
if $c_0 = 0$, if $c_1,\dots ,c_n\ge 1$, then
\begin{equation}\label{decompositionRotation}
\sum_{k=0}^{c_1 +\dots + c_n - 1}\mathrm R_{k\alpha }u = \sum_{j=0}^n \mathrm R_{(c_0 + \dots + c_{j-1})\alpha }\sum_{k=0}^{c_j - 1}\mathrm R_{k\alpha }u.
\end{equation}

\noindent\emph{Proof of Proposition \ref{result1}.} 
By point 4) of Lemma \ref{fraccont}, there exists a subsequence $(\tilde p_k/\tilde q_k)_{k\ge 1}\subset (p_n/q_n)_{n\ge 0}$ such that $|\alpha - \tilde p_k/\tilde q_k |<1/2\tilde q_k^2$. 
Moreover, once $\tilde q_1,\dots ,\tilde q_k$ $(k\ge 1)$ are given, one may take $\tilde q_{k+1}$ as large as we please (by still taking a subsequence).
So, by Lemma \ref{Denjoy}, $\mathrm R_\beta y_{\tilde q_k }$ takes only the values $\pm 1$ ($\beta\in\mathbf R$).
For $k\ge 1$, let $n_k = \tilde q_1 + \dots + \tilde q_k$  and define  $f_1 = y_{\tilde q_1}$ and $f_k = \mathrm R_{n_{k-1}\alpha}y_{\tilde q_k}$
(thus $f_k(x)=\pm 1$ and $f_k^2(x)=1$ for every $x\in\mathbf T$ and every $k\ge 1$).
By (\ref{decompositionRotation}), one has $y_{n_k}=\sum_{j=1}^k f_j$. 

Let $(\delta_k)_{k\ge 1}\subset\mathbf R_0^+$ be such that $\sum_{j=1}^k\delta_j/\sqrt k\rightarrow 0$ as $k\rightarrow\infty$.
One may suppose that, for every $k\ge 1$, and for every $\gamma\in\lbrack -1,1\rbrack$
\begin{equation}\label{decooo1}
|\int_0^1 f_{n_{k+1}}e^{i\gamma(f_{n_1}+\dots +f_{n_k})}\mathrm d x  | \le   \delta_k.
\end{equation}
Indeed, for some $m(k)\in\mathbf N$,
one may write $\lbrack 0,1\rbrack= \bigcup_{j=1}^{m(k)}\mathrm I_j$, 
in such a way that $e^{i\gamma(f_{n_1}+\dots +f_{n_k})}$ is constant on each $\mathrm I_j$ ($1\le j\le m(k)$).
But, by Lemma \ref{faible}, one may suppose that $|\int_{\mathrm I_j}f_{n_{k+1}}\mathrm d x|\le\delta_k/m(k)$ ($1\le j\le m(k)$) ; 
indeed one just needed to take $\tilde q_{k+1}$ large enough.

Let $\lambda\in\mathbf R$. 
For $k\ge 2$ and for $2\le j\le k$, one has
\begin{equation}\label{aiterer}
e^{i\frac{\lambda}{\sqrt k }(f_{n_1}+\dots +f_{n_j})} = e^{i\frac{\lambda}{\sqrt k }(f_{n_1}+\dots +f_{n_{j-1}})}
(1+i\frac{\lambda  }{\sqrt k } f_{n_j} - \frac{\lambda^2}{2k} + \mathcal O\frac{ |\lambda|^3}{k^{3/2}}).
\end{equation}
For $k\ge 1$ big enough, one has $|\lambda /\sqrt k|\le 1$ and $|1 - \frac{\lambda^2}{2k}|\le 1$.
Therefore, using (\ref{aiterer}) recursively, and applying (\ref{decooo1}), one finds that
\[
|\int_0^1 e^{i\frac{\lambda}{\sqrt k}(f_{n_1}+\dots +f_{n_k})}\mathrm d x - (1-\frac{\lambda^2}{2k})^{k}|\le
\frac{|\lambda |}{\sqrt k}\sum_{j=1}^{k-1} \delta_j + \mathcal O\frac{|\lambda |^3}{\sqrt k}. 
\]
So, for each $\lambda\in\mathbf R$, $\int_0^1 e^{i\frac{\lambda }{\sqrt k }y_{n_k} }\mathrm d x \rightarrow e^{-\lambda^2/2 }$ as $k\rightarrow\infty$. $\square$

We now give an heuristic link between Proposition \ref{result1} and the theory of expanding maps of the interval \cite{hof}.
If $k\ge 2$ is an integer, one defines the map $\mathrm T_k : \mathbf T\rightarrow\mathbf T, x\mapsto \overline{kx}$ 
(with the notation $\overline x = x - \lfloor x \rfloor$). 
For $n\ge 1$, $\mathrm T_k^n = \mathrm T_k \circ \dots \circ \mathrm T_k = \mathrm T_{k^n}$.

Let $\psi_*$ be the function given by (\ref{psistar}). If $k\ge 2$, $(\psi_*\circ\mathrm T_k^n)_{n\ge 1 }$ is a sequence of random variables on $(\mathbf T, m_\mathrm L)$. 
One shows that there exists $\sigma_k>0$ such that
\begin{equation}\label{comppp}
\frac{1}{\sqrt n }\sum_{j=1}^n \psi_*\circ \mathrm T_k^j = \frac{1}{\sqrt n } \sum_{j=1}^n \psi_*\circ \mathrm T_{k^j}  \stackrel{\mathrm D}{\rightarrow}\mathrm G(\sigma_k)\quad \mathrm{as}\quad n\rightarrow \infty.
\end{equation}
Indeed, if $k$ is even, the random variables $(\psi_*\circ \mathrm T_k^n )_{n\ge 1 }$ are actually independent and equidistributed (so $\sigma_k = 1$).
In general, one may use Theorem 5 of \cite{hof}  : one checks that $\mathrm T_k$ is mixing with respect to the invariant measure $m_\mathrm L$, 
and that the equation $u\circ\mathrm T_k - u = \psi_*$ admits no solution $u\in\mathrm L^2(\mathbf T,\mathbf R)$ (by Fourier expansion for example), so that $\sigma_k>0$.

Let us now consider the sequence $(f_k)_{k\ge 1 }$ constructed in the proof of Proposition \ref{result1} (we keep the notations of this part). 
One has $f_k = \mathrm R_{n_{k-1}\alpha }y_{\tilde q_k}$ $(k\ge 2)$. 
First, one may expect the rotation $\mathrm R_{n_{k-1}\alpha}$ to play no essential role in the decorrelation properties of the variables $f_k$ $(k\ge 1)$.
Next, the proof of Lemma \ref{Denjoy} was entirely based on the fact that $y_{\tilde q_k }$ may be approximated by $\psi_*\circ\mathrm T_{\tilde q_k}$. 
For each irrational number $\alpha$, the sequence $(q_k)_{k\ge 0 }$ grows at least exponentially with $k$ (a superexponential growth improves actually the decorrelations). 

One comes thus to the conclusion that the sequence $y_{n_k}/\sqrt k = \sum_{j=1}^k f_j/\sqrt k$ is likely to have a statistical behavior analogous to (\ref{comppp}).
The proof of Proposition \ref{result1} was greatly simplified by the fact that one allowed $\tilde q_k$ to grow arbitrarily fast with $k$. 
In the two next Sections, we prove basically that an exponential growth is enough in some cases.

\section{Central Limit Theorem}\label{sec3}

Let $\mu$ be a probability measure on $\mathbf T$. 
In this Section, $\mathrm L^p(\mathbf T, \mathbf R)=\mathrm L^p(\mathbf T, \mathbf R,\mathrm d \mu)$ ($p\ge 1$).
\begin{Proposition}\label{clt}
Let $(q_k)_{k\ge 1}\subset \mathbf N_0$, and suppose there exists $\mathrm \rho>0$ such that for every $k\ge 1$,
\begin{equation}\label{expcrois} 
q_{k+1}\ge e^{2\rho }q_k.
\end{equation} 
Let $(f_{jk})_{j,k\ge 1}\subset \mathrm{BV}(\mathbf T,\mathbf R)$ 
be  random variables on $(\mathbf T,\mu)$ such that $\int_0^1 f_{jk}\mathrm d \mu =0$ $(j,k\ge 1)$. 
Let $\mathrm S_n = f_{n1}+\dots +f_{nn}$.
Suppose that there exists $\mathrm C>0$ such that 

$1)$ for every $j,k\ge 1$, 
$||f_{jk}||_{\mathrm L^\infty}\le \mathrm C$ and $\mathrm{Var}(f_{jk})\le \mathrm C q_k$, 

$2)$ for some $\beta\in\mathbf R$, for every $\phi\in\mathrm{BV}(\mathbf T,\mathbf R)$ such that $\int_0^1\phi\mathrm d \mu =0$, for $j\ge 1$ and for $t\ge s \ge 1$, 
\begin{eqnarray}
|\int_0^1 \phi .  f_{js}\mathrm d\mu | &\le& \mathrm C. \mathrm{Var}(\phi)\frac{s^\beta}{q_s}, \label{decor1}\\ 
|\int_0^1 \phi .  f_{js} f_{jt}\mathrm d\mu | &\le & \mathrm C. \mathrm{Var}(\phi)\frac{s^\beta}{q_s},\label{decor2}
\end{eqnarray}

$3)$ if $\sigma_n = ||\mathrm S_n/\sqrt n||_{\mathrm L^2}$, there exists $\epsilon >0$ such that $\epsilon \le \sigma_n \le \mathrm C$ for each $n\ge 1$.\\
Then, $\mathrm S_n/\sqrt n\sigma_n \stackrel{\mathrm D}{\rightarrow}\mathrm G(1)$ as $n\rightarrow \infty$.
\end{Proposition}

\noindent\emph{Proof.}  
To simplify notations, 
one will only consider the case where $f_{jk}=f_{lk}$ for all $j,l,k\ge 1$, and write $f_k = f_{1k}$ ($k\ge 1$). 
The hypotheses of the Proposition only involve estimates which are independent of the index $j$ of $f_{jk}$ ($j,k\ge 1$).
Therefore, the proof of the general case is a straightforward adaptation of this one. 
We begin by a Lemma.
\begin{Lemma}\label{lfour}
Under the hypothesis of Proposition \ref{clt}, there exists $\mathrm C >0$ such that for every $m,n\ge 1$, 
$||\sum_{k=m}^{m+n}f_k||_{\mathrm L^4}\le \mathrm C (n\ln (n+m))^{1/2}$.
\end{Lemma}
\emph{Proof.} One has 
\begin{equation}\label{dedef}
||\sum_{k=m}^{m+n}f_k||_{\mathrm L^4}^4 = \sum_{m\le s,t,u,v\le  m+n}\int_0^1 f_s f_t f_u f_v\mathrm d \mu  \le
4! \sum_{m \le s\le t\le u\le v\le m+n }|\int_0^1 f_s f_t f_u f_v\mathrm d \mu | .
\end{equation}
Until the end of this proof, one assumes $m \le s\le t\le u\le v\le m+n$.
One defines
\begin{equation}\label{dededef}
\mathrm S_{stuv} = |\int_0^1 f_s f_t f_u f_v\mathrm d \mu |
\quad\mathrm{and}\quad
\mathrm S_{tu} = \sum_{s,v}\mathrm S_{stuv}.
\end{equation}

By hypothesis 1), one has always $\mathrm S_{stuv} \le \mathrm C$. 
Let us obtain two others estimates of this quantity.
If $f,g\in\mathrm{BV}(\mathbf T,\mathbf R)$, one has $\mathrm{Var}(fg)\le ||f||_{\mathrm L^\infty}\mathrm{Var}(g)+||g||_{\mathrm L^\infty}\mathrm{Var}(f)$. 
Therefore, by 1), one has
\[
\mathrm{Var}(f_s f_t f_u)\le \mathrm C (q_s + q_t + q_u)\le 3\mathrm C q_u.
\]
By (\ref{decor1}) and (\ref{expcrois}), one first obtains that
\begin{equation}\label{i1}
\mathrm S_{stuv} =  |\int_0^1 (f_s f_t f_u). f_v\mathrm d\mu | \le  \mathrm C \frac{q_u}{q_v}v^\beta \le\mathrm C e^{-2\rho(v-u)}v^\beta.
\end{equation}
Similarly, by (\ref{decor2}), (\ref {decor1}) and (\ref{expcrois}), one then gets
\begin{eqnarray}
\mathrm S_{stuv}
& \le & 
|\int_0^1 \Big(   f_sf_t - \int_0^1 f_sf_t\mathrm d\mu  \Big) . f_u f_v \mathrm d \mu |
+ |\int_0^1 f_s . f_t \mathrm d \mu | . |\int_0^1 f_u f_v \mathrm d \mu | \nonumber\\
& \le &
\mathrm C (\frac{q_t}{q_u}u^\beta + \frac{q_s}{q_t}t^\beta)\le 
\mathrm C(e^{-2\rho(u-t)}u^\beta + e^{-2\rho(t-s)}t^\beta). \label{i2}
\end{eqnarray}
Set $\kappa = (\beta/\rho)\ln (n+m)$. One assumes $\kappa\ge 1$.
Inequalities (\ref{i1}) and (\ref{i2}) imply respectively
\begin{eqnarray}
\mathrm S_{stuv} &\le &\mathrm C e^{-\rho(v-u)} 
\quad\mathrm{if} \quad 
v-u\ge \kappa , \label{locloc1}\\
\mathrm S_{stuv} &\le & \mathrm C (e^{-\rho(u-t)} + e^{-\rho (t-s)}) 
\quad\mathrm{if} \quad
u-t\ge \kappa\quad\mathrm{and} \quad t-s\ge \kappa .\label{locloc2}
\end{eqnarray}

We now estimate $\mathrm S_{tu}$ in (\ref{dedef}) for fixed $t,u$. First we consider the case $u-t < \kappa$.
By (\ref{locloc1}) one gets (setting $v-u = k$ after the second inequality)
\begin{eqnarray}
\mathrm S_{tu} 
&=&
\sum_{s,v:v-u<\kappa }\mathrm S_{stuv} + \sum_{s,v:v-u\ge\kappa }\mathrm S_{stuv} \nonumber \\
&\le&
\sum_{s,v:v-u<\kappa }\mathrm C +  \sum_{s,v:v-u\ge\kappa }\mathrm C e^{-\rho(v-u)}
\le
\sum_{k<\kappa,s }\mathrm C + \sum_{k\ge \kappa,s }\mathrm C e^{-\rho k }
\le
\mathrm C n\kappa + \mathrm C n \le \mathrm C n\kappa .\label{brolleke1}
\end{eqnarray}

Next, we consider the case $u-t \ge \kappa$. 
We write the decomposition $\mathrm S_{tu}=\mathrm S_{tu}(1)+\mathrm S_{tu}(2)+\mathrm S_{tu}(3)$. 
Those three terms will be defined one by one.
First, in the same way as (\ref{brolleke1}), one gets
\begin{equation}\label{brolleke2}
\mathrm S_{tu}(1) \stackrel{\Delta }{=}
\sum_{s,v : u-t \le v-u }\mathrm S_{stuv} \le
\sum_{s,v : u-t \le v-u }\mathrm C e^{-\rho(v-u) }\le \mathrm C n e^{-\rho (u-t) }.
\end{equation}
Then, if $v-u < u-t \le t-s$, one uses (\ref{locloc2}) to get $\mathrm S_{stuv}\le \mathrm C e^{-\rho (u-t) }$, and so 
\begin{equation}\label{brolleke3}
\mathrm S_{tu}(2) 
\stackrel{\Delta }{=}
\sum_{s,v : v-u<u-t\le t-s }\mathrm S_{stuv} 
\le
\mathrm C e^{-\rho (u-t) }\sum_{k<u-t,s}1
\le
\mathrm C n (u-t)e^{-\rho (u-t) }. 
\end{equation}
Finally, let $\mathrm B = \lbrace (s,v) : v-u < u-t,  t-s < u-t \rbrace$. 
For at most $\kappa^2$ elements $(s,v)\in\mathrm B$, one has $v-u <\kappa$ and $t-s <\kappa$, and so only the estimate $\mathrm S_{stuv}\le \mathrm C$. 
For all the others, one has
$\mathrm S_{stuv}\le \min\lbrace \mathrm C e^{-\rho (v-u) },\mathrm C e^{-\rho (t-s) }\rbrace$
by (\ref{locloc1}) and (\ref{locloc2}) (to use (\ref{locloc2}) one uses the fact that $t-s < u-t$). Therefore, 
\begin{equation}\label{brolleke4}
\mathrm S_{tu}(3) = \sum_{s,v\in\mathrm B }\mathrm S_{stuv} \le \mathrm C\kappa^2 + \mathrm C \sum_{j,k\ge 0 }\min \lbrace e^{-\rho j }, e^{-\rho k } \rbrace \le \mathrm C\kappa^2 + \mathrm C \le \mathrm C\kappa^2 .
\end{equation}

By (\ref{dedef} - \ref{dededef}) and (\ref{brolleke1} - \ref{brolleke4}) one obtains
\[
||\sum_{k=m}^{m+n}f_k||_{\mathrm L^4}^4 
\le
\mathrm C \sum_{t,u : u-t<\kappa } n\kappa + 
\mathrm C \sum_{t,u} \Big(
ne^{-\rho (u-t) } + n(u-t)e^{-\rho (u-t) } + \kappa^2
\Big) .
\]
Because $\kappa = (\beta/\rho)\ln (n+m)$, this gives the result. $\square$

Let us now come to the proof of the Proposition. 
Like in \cite{bun}, one defines a kind of coarse grained variables that get more and more decorrelated as $n\rightarrow\infty$.
Let $\gamma_1,\gamma_2\in\rbrack 0,1\lbrack$ be such that $\gamma_1 > \gamma_2 + 1/2$ (and thus $\gamma_2 <1/2$).
If $n\ge 1$, set $n_1 = \lfloor n^{\gamma_1 }\rfloor$ and $n_2 = \lfloor n^{\gamma_2 }\rfloor$. 
In the sequel, we suppose that $n$ is large enough to have $n_2\ge 1$.
One writes $\mathrm S_n = \sum_{k=1}^{p(n)}(\mathrm X_{nk}+\mathrm Y_{nk})$ 
where $p(n)$ is the smallest integer such that $p(n).(n_1 +n_2)\ge n$ and where
\begin{eqnarray}
\mathrm X_{nk} &=& f_{(k-1)(n_1+n_2)+1} +\dots + f_{(k-1)(n_1 + n_2) +n_1},\label{Xcoarse} \\
\mathrm Y_{nk} &=& f_{(k-1)(n_1 + n_2) + n_1 + 1} + \dots + f_{k(n_1+n_2)}\label{Ycoarse}
\end{eqnarray}
($1\le k\le p(n)-1$ ; for $k=p(n)$ the definition is the same but one puts $0$ instead of $f_j$ whenever $j>n$). 
One has $p(n)/n^{1-\gamma_1 }\rightarrow 1$ as $n\rightarrow\infty$.

Let $\lambda\in\mathbf R$. This number will be treated as a constant in all our estimates.
For $n\ge 1$, one defines 
\begin{equation}\label{caract1}
\mathrm J_n(\lambda)=\int_0^1e^{i\frac{\lambda}{\sqrt n\sigma_n }\mathrm S_n}\mathrm d \mu, 
\end{equation}
and, for $1\le k\le p(n)$, 
\begin{equation}\label{caract2}
\mathrm I_{nk}(\lambda)=\int_0^1 e^{i\frac{\lambda}{\sqrt n\sigma_n }(\mathrm X_{n1}+\dots +\mathrm X_{nk})}\mathrm d \mu. 
\end{equation}
One puts also $\mathrm I_{n0}(\lambda)=1$.
By hypothesis 3) and (\ref{Ycoarse}), one has
\begin{equation}\label{ine1}
|\mathrm J_n(\lambda) - \mathrm I_{np(n)}(\lambda)|
\le ||\frac{\lambda}{\sqrt n\sigma_n }\sum_{k=1}^{p(n)}\mathrm Y_{nk}||_{\mathrm L^\infty}
\le \mathrm C  \frac{p(n)n_2}{\sqrt n} \le \mathrm C n^{-(\gamma_1 - (\gamma_2 +1/2))}.
\end{equation}

\begin{Lemma}\label{caracteristic}
Under the hypothesis of Proposition \ref{clt} and if $1\le k\le p(n)$, one has
\[
\mathrm I_{nk}(\lambda)=(1-\frac{\lambda^2}{2n\sigma_n^2 }\int_0^1\mathrm X_{nk}^2\mathrm d\mu)\mathrm I_{n(k-1)}(\lambda) + r_{nk}(\lambda)
\]
with 
$|r_{nk}(\lambda)|\le \mathrm C (n^{\beta +1}e^{-2\rho n_2 } + n^{-\frac{3}{2}(1-\gamma_1) }\ln^{3/2}n)$ (where $\mathrm C$ depends on $\lambda$).
\end{Lemma}
\emph{Proof.} Let us only consider the most difficult case $k\ge 2$.
To simplify formulas, one will assume that $\sigma_n = 1$ for all $n\ge 1$. By hypothesis 3), this does not change our estimates. 
One has
\begin{equation}\label{tayloor}
\mathrm I_{nk}(\lambda)=\int_0^1
(1 + i\frac{\lambda}{\sqrt n}\mathrm X_{nk} - \frac{\lambda^2}{2n}\mathrm X_{nk}^2  + \mathcal O\frac{|\mathrm X_{nk} |^3}{n^{3/2}})
e^{i\frac{\lambda}{\sqrt n}(\mathrm X_{n1}+\dots +\mathrm X_{n(k-1)})}\mathrm d \mu.
\end{equation}
If $g\in\mathcal C^1(\mathbf R,\mathbf R)\cap\mathrm L^\infty(\mathbf R,\mathbf R)$, and if $u\in\mathrm{BV}(\mathbf T,\mathbf R)$, 
then $\mathrm{Var}(g\circ u)\le ||g'||_{\mathrm L^\infty }\mathrm{Var}(u)$.  
By (\ref{expcrois}), one has $q_1 + \dots + q_n \le \mathrm C q_n$.
So, by (\ref{Xcoarse}) and hypothesis 1), one has
\begin{equation}\label{varvar}
\mathrm{Var}(e^{i\frac{\lambda}{\sqrt n}(\mathrm X_{n1}+\dots +\mathrm X_{n(k-1)})}) \le \frac{\mathrm C}{\sqrt n}q_{(k-2)(n_1+n_2)+n_1}.
\end{equation}
Therefore, first, 
using (\ref{Xcoarse}), (\ref{decor1}), (\ref{varvar}) and (\ref{expcrois}), 
one gets
\begin{eqnarray}
|\int_0^1 \frac{\lambda}{\sqrt n}\mathrm X_{nk}e^{i\frac{\lambda}{\sqrt n}(\mathrm X_{n1}+\dots +\mathrm X_{n(k-1)})}\mathrm d\mu | &\le &
\frac{\lambda}{\sqrt n}\sum_{j=1}^{n_1}|\int_0^1 e^{i\frac{\lambda}{\sqrt n}(\mathrm X_{n1}+\dots +\mathrm X_{n(k-1)})}. f_{(k-1)(n_1+n_2)+j}\mathrm d\mu | \nonumber\\
& \le &
 \frac{\mathrm C }{n}\sum_{j=1}^{n_1}\frac{q_{(k-2)(n_1+n_2)+n_1}}{q_{(k-1)(n_1+n_2)+j}}((k-1)(n_1+n_2)+j)^\beta \nonumber\\
&\le & \frac{\mathrm C}{n}n_1 e^{-2\rho n_2}n^\beta 
\le \mathrm C n^{\beta}e^{-2\rho n_2}. \label{inoue1}
\end{eqnarray}
Then, similarly, using (\ref{decor2}) instead of (\ref{decor1}), and noticing that 
nor $\mathrm X_{nk}^2$ nor $e^{i\frac{\lambda}{\sqrt n}(\mathrm X_{n1}+\dots +\mathrm X_{n(k-1)})}$ have in general a zero integral, one gets
\begin{equation}\label{inoue2}
|\int_0^1 \frac{\lambda^2}{2n}\mathrm X_{nk}^2e^{i\frac{\lambda}{\sqrt n}(\mathrm X_{n1}+\dots +\mathrm X_{n(k-1)})}\mathrm d \mu - 
\frac{\lambda^2}{2n}\int_0^1\mathrm X_{nk}^2\mathrm d \mu . 
\int_0^1e^{i\frac{\lambda}{\sqrt n}(\mathrm X_{n1}+\dots +\mathrm X_{n(k-1)})}\mathrm d\mu |\le \mathrm C n^{\beta+1} e^{-2\rho n_2}. 
\end{equation}
Finally, by Lemma \ref{lfour} and (\ref{Xcoarse}), one has
\begin{eqnarray}
|\int_0^1 
 \frac{|\mathrm X_{nk} |^3}{n^{3/2}}
e^{i\frac{\lambda}{\sqrt n}(\mathrm X_{n1}+\dots +\mathrm X_{n(k-1)})}\mathrm d \mu |   &\le & 
\frac{\mathrm C }{n^{3/2}}||\mathrm X_{nk}||_{\mathrm L^3}^3\le \frac{\mathrm C}{n^{3/2}}||\mathrm X_{nk}||_{\mathrm L^4}^3 
\le \frac{\mathrm C }{n^{3/2}}(n_1\ln n)^{3/2} \nonumber \\
&\le& \mathrm C n^{-\frac{3}{2}(1-\gamma_1) }\ln^{3/2}n.\label{inoue3}
\end{eqnarray}

Inserting (\ref{inoue1}-\ref{inoue3}) in (\ref{tayloor}), one gets the result. $\square$

To prove Proposition \ref{clt}, it is enough to show that $\mathrm J_n(\lambda)\rightarrow e^{-\lambda^2/2 }$ as $n\rightarrow\infty$. 
For $n$ large enough, one has $|1-(\lambda^2/2n)\int_0^1\mathrm X_{nk}^2\mathrm d x |\le 1$. Thus, by (\ref{ine1}) and by recursive application of Lemma \ref{caracteristic}, one has 
\[
|\mathrm J_n(\lambda) - (1-\frac{\lambda^2}{2n\sigma_n^2 }\int_0^1\mathrm X_{np(n)}^2\mathrm d \mu)\dots (1-\frac{\lambda^2}{2n\sigma_n^2 }\int_0^1\mathrm X_{n1}^2\mathrm d \mu) |\le 
\]\[
\mathrm C p(n)(n^{\beta +1}e^{-2\rho n_2 } + n^{-\frac{3}{2}(1-\gamma_1) }\ln^{3/2}n) + \mathrm C n^{-(\gamma_1 - (\gamma_2 +1/2))}.
\]
Because $p(n)/n^{1-\gamma_1 }\rightarrow 1$ as $n\rightarrow\infty$, the right hand side of this inequality goes to 0 as $n\rightarrow\infty$.

Therefore, it is enough to show that
$\ln\prod_{k=1}^{p(n)}(1-(\lambda^2/2n)\int_0^1\mathrm X_{nk}\mathrm d\mu )\rightarrow -\lambda^2/2$ as $n\rightarrow\infty$.
By hypothesis 3), one has
\[
\ln \prod_{k=1}^{p(n)}(1-\frac{\lambda^2}{2n\sigma_n^2 }\int_0^1\mathrm X_{nk}^2\mathrm d\mu) = 
-\frac{\lambda^2}{2n\sigma_n^2 }\sum_{k=1}^{p(n)}\int_0^1\mathrm X_{nk}^2\mathrm d \mu+ 
\mathcal O \frac{1}{n^2}\sum_{k=1}^{p(n)}(\int_0^1\mathrm X_{nk}^2\mathrm d\mu)^2 .
\]
First, Lemma \ref{lfour} is still valid if  $||.||_{\mathrm L^2}$ is used in place of $||.||_{\mathrm L^4}$ (because $||.||_{\mathrm L^2}\le ||.||_{\mathrm L^4}$), and so, by (\ref{Xcoarse}),
\[
\frac{1}{n^2}\sum_{k=1}^{p(n)}(\int_0^1\mathrm X_{nk}^2\mathrm d\mu)^2 
=\frac{1}{n^2}\sum_{k=1}^{p(n)}||\mathrm X_{nk} ||^4_{\mathrm L^2 }
\le \frac{\mathrm C }{n^2}p(n)n_1^2 \ln^2 n
\le \frac{\mathrm C }{n^2}n^{1-\gamma_1 }n^{2\gamma_1 }\ln^2 n
\rightarrow 0 
\quad\mathrm{as}\quad n\rightarrow\infty . 
\]
Next, one has
\[
\int_0^1 (\sum_{k=1}^{p(n)}\mathrm X_{nk} )^2\mathrm d\mu = 
\int_0^1 \sum_{k=1}^{p(n)}\mathrm X_{nk}^2\mathrm d\mu + 2 \sum_{j=2}^{p(n)}\sum_{k=1}^{j-1}\int_0^1\mathrm X_{nj}\mathrm X_{nk}\mathrm d\mu.
\]
By (\ref{Xcoarse}) and hypothesis 1), 
$\mathrm{Var}(\mathrm X_{nk})\le \mathrm C q_{(k-1)(n_1 + n_2)+n_1}$.
So, by (\ref{Xcoarse}), (\ref{decor1}) and (\ref{expcrois}), one obtains as for (\ref{inoue1}) that, if $1\le k < j\le p(n)$, 
\[
|\int_0^1\mathrm X_{nj}\mathrm X_{nk}\mathrm d\mu | \le \sum_{l=1}^{n_1}|\int_0^1 \mathrm X_{nk}f_{(j-1)(n_1+n_2)+n_1} \mathrm d\mu | \le \mathrm C n^\beta e^{-2\rho n_2 }.
\]
Therefore
\[
\int_0^1\sum_{k=1}^{p(n)}\mathrm X_{nk}^2\mathrm d \mu-\int_0^1 (\sum_{k=1}^{p(n)}\mathrm X_{nk})^2\mathrm d \mu\rightarrow 0
\quad\mathrm{as}\quad n\rightarrow\infty. 
\]
By (\ref{Ycoarse}), 
\[
||\frac{1}{\sqrt n}\sum_{k=1}^{p(n)}\mathrm Y_{nk}||_{\mathrm L^2 } \le ||\frac{1}{\sqrt n }\sum_{k=1}^{p(n)}\mathrm Y_{nk}||_{\mathrm L^\infty }\rightarrow 0
\quad\mathrm{as}\quad n\rightarrow\infty. 
\]
Therefore $\sum_{k=1}^{p(n)}\int_0^1 \mathrm X_{nk}^2\mathrm d\mu/n\sigma_n^2 - ||\mathrm S_n ||^2_{\mathrm L^2 }/n\sigma_n^2\rightarrow 0$ as $n\rightarrow\infty$.
One concludes by hypothesis 3). $\square$

\section{Proof of Proposition \ref{result3} and Corollary \ref{result2}}\label{sec4}

\emph{Proof of Proposition \ref{result3}.}
Let $\alpha$ be a number of constant type (see (\ref{constanttype})), 
and let $d$ be the constant given by (\ref{ead}).  
Let $(p_n/q_n)_{n\ge 0 }$ be its convergents, and $(a_n)_{n\ge 0 }$ be its partial quotients. 
One may decompose an integer $r\ge 1$ according to the \emph{Ostrowski system of numeration}.
One has $q_n\le r < q_{n+1}$ for some $n\ge 0$. One writes
\begin{equation}\label{canonicalDecomposition}
r = b_n q_n  + \dots + b_0 q_0,
\end{equation}
where $b_k\ge 0$ ($0\le k\le n$) are integers defined recursively : 
first $b_n\ge 1$ is the only integer such that $0\le r-b_nq_n < q_n$ ; 
then, if $b_{k+1}\ge 0$ ($0\le k\le n-1$) is given, $b_k\ge 0$ is the only integer such that $0 \le r - (b_nq_n + \dots +b_kq_k)<q_k$. 
One has $b_k\le a_{k+1}$, and by (\ref{ead}), $b_k\le d$ ($1\le k\le n$). 

Now let us consider a sequence $(s_n)_{n\ge 1}$ such that $q_n\le s_n < q_{n+1}$.
One wants to show that $z_{s_n}/||z_{s_n}||_{\mathrm L^2}\stackrel{\mathrm D }{\rightarrow}\mathrm G(1)$ as $n\rightarrow\infty$. 
All our estimates will be independent of the choice of the sequence $(s_n)_{n\ge 1 }$. 
Therefore, 
this will actually imply that 
$z_{n}/||z_{n}||_{\mathrm L^2}\stackrel{\mathrm D }{\rightarrow}\mathrm G(1)$ as $n\rightarrow\infty$. 
Indeed, for $n\ge q_1$, 
one finds a sequence $(s_k)_{k\ge 1 }$ with $q_k \le s_k<q_{k+1}$ for every $k$, and
such that $n=s_j$ for some $j\ge 1$ ; one has
$j\rightarrow\infty$ as $n\rightarrow \infty$.

For each $n\ge 1$, there exists one and only one $1 \le r_n\le s_n$ such that $y_{r_n}=z_{s_n}$. 
Let $(b_k(n))_{1\le k\le n }$ be the sequence associated to the canonical decomposition (\ref{canonicalDecomposition}) of $r_n$ 
($b_k(n)$ may not be defined for large $k$, one set then $b_k(n)=0$).
Let us define $(f_{jk})_{j,k\ge 1 }\subset\mathrm{BV}(\mathbf T,\mathbf R)$. 
Let $j\ge 1$. 
If $1\le k\le j$, one sets 
\begin{equation}\label{froulouk}
f_{jk} = \mathrm R_{(b_0(j)q_0 + \dots + b_{k-1}(j)q_{k-1})\alpha }\sum_{l=0}^{b_k(j)q_k - 1}\mathrm R_{ l \alpha }\psi.
\end{equation}
If $k>j$, one sets $f_{jk}=0$. 
By (\ref{decompositionRotation}), one has $z_{s_n} = y_{r_n} = \sum_{1\le k\le n }f_{nk}$.

Therefore, it is enough to show that the sequences $(q_k)_{k\ge 1 }$ and $(f_{jk})_{jk\ge 1 }$ satisfy the hypotheses of Proposition \ref{clt}.
The estimates on $||z_n||_{\mathrm L^2 }$ will then be also proven. 
Indeed, we will need to show that hypothesis 3) is satisfied, and thus that $\epsilon \sqrt n \le ||z_{s_n}||_{\mathrm L^2 }\le \mathrm C\sqrt n$ (where $q_n\le s_n < q_{n+1}$). 

First, (\ref{expcrois}) is satisfied.
One has $q_2>q_1$, and, for $n\ge 2$, $q_{n-1}\ge q_n/2d$ by (\ref{ead}). Therefore, for $n\ge 2$,
\[ 
q_{n+1}\ge q_n + q_{n-1}\ge (1 + 1/2d)q_n. 
\]
Next, hypothesis 1) is satisfied. 
By (\ref{froulouk}), (\ref{decompositionRotation}), (\ref{ead}) and Denjoy-Koksma inequality (\ref{DenjoyKoksma}), one has for all $j,k\ge 1$
\[
||f_{jk}||_{\mathrm L^\infty }\le b_k(j)||\sum_{l=0}^{q_k-1}\mathrm R_{l\alpha }\psi ||_{\mathrm L^\infty }\le d.\mathrm{Var}(\psi). 
\]
And, 
by (\ref{froulouk}) and (\ref{ead}), one has, for all $j,k\ge 1$
\[
\mathrm{Var}(f_{jk})\le \mathrm{Var}(\psi)b_k(j)q_k\le d.\mathrm{Var}(\psi)q_k.  
\]

Let us show that hypothesis 2) is satisfied. 
\begin{Lemma}\label{puissance}
If $\alpha$ is of constant type,
then 
$\forall p\in\mathbf N$, $\exists c_0\in\mathbf N$ $:$ $(\forall c\in\mathbf N : c\ge c_0)$, $\forall n\in\mathbf N$, $q_{cn}\ge q_n^p$.
\end{Lemma}
\emph{Proof.}
Let $d$ be given by (\ref{ead}). For $c$ large enough and $n\ge 1$, one has by (\ref{expcrois}),
$q_{cn}\ge e^{2\rho cn}\ge (2d)^{pn}\ge q_n^p$. $\square$

Like in \cite{bri}, one defines  the sets 
\begin{equation}\label{gamman}
\Gamma_1 = \lbrace j\in\mathbf N_0 : 1/q_1 \le |1-e^{2i\pi j\alpha}|\rbrace , \qquad
\Gamma_n = \lbrace j\in\mathbf N_0 : 1/q_n \le |1-e^{2i\pi j\alpha}| < 1/q_{n-1}\rbrace \quad (n\ge 2). 
\end{equation}
One has $\mathbf N_0 = \bigcup_{n\ge 1}\Gamma_n$ and $\Gamma_m\cap \Gamma_n = \emptyset$ if $m\ne n$.
\begin{Lemma}
There exists $\mathrm C > 0$ such that, for every $m\ge 1$,
\begin{equation}\label{latticeinequality}
j\in\Gamma_m \Rightarrow j\ge \mathrm C q_m 
\quad\mathrm{and}\quad 
j,k \in\bigcup_{n\ge m}\Gamma_n \Rightarrow |k-j|\ge \mathrm Cq_m.
\end{equation} 
\end{Lemma}
\emph{Proof.}
If $j\in\Gamma_m$, $|1-e^{2i\pi j\alpha }|<1/q_{m-1}$. 
By (\ref{constanttype}) and (\ref{normequ}), there exists $\mathrm C> 0$ such that 
$4\mathrm C/|j|\le 4|j\alpha |_\mathbf T \le 1/q_{n-1}$. 
Let $d$ be given by (\ref{ead}). One has 
$q_{m-1}\ge q_m/2d$, and so $|j|\ge \mathrm C q_m$. 
Next, if $j,k \in\bigcup_{n\ge m}\Gamma_n$, one has by (\ref{triangle}) that
\[
|1 - e^{2i\pi (k-j)\alpha }| \le |1 - e^{2i\pi k\alpha }| + |1-e^{2i\pi j\alpha }|\le \frac{2}{q_{m-1}},  
\]
and so one gets the second inequality (\ref{latticeinequality}) as the first. $\square$

\begin{Lemma}\label{decorel}
There exist $\mathrm C,\beta\in\mathbf R$ such that, for every $\phi\in\mathrm{BV}(\mathbf T,\mathbf R)$ with $\int_0^1\phi\mathrm d x =0$, for $l\ge 1$ and for $t\ge s \ge 1$, 
\begin{equation}\label{blla}
|\int_0^1 \phi  f_{ls}\mathrm dx|\le\mathrm C. \mathrm{Var}(\phi)\frac{s^\beta}{q_s}, \qquad
|\int_0^1 \phi  f_{ls} f_{lt}\mathrm dx|\le\mathrm C. \mathrm{Var}(\phi)\frac{s^\beta}{q_s}.
\end{equation}
\end{Lemma}
\emph{Proof.}
Both inequalities may be shown in the same way, but the first one is simpler, and we will only prove the second one. 
In this proof, we consider as constants, numbers that depend only on $\alpha$ or $\psi$.
Let $l\ge 1$ and $t\ge s \ge 1$. One writes 
\[
g=f_{ls}\quad\mathrm{and}\quad h=f_{lt}.
\] 
Let $\phi\in\mathrm{BV}(\mathbf T,\mathbf R)$. One has $|\hat\phi (k)|\le \mathrm{Var}(\phi)/2\pi|k|$ if $k\in\mathbf Z_0$, and $\hat\phi(0)=0$ by hypothesis.
Let us first simplify the problem in two ways. 

First,
by Lemma \ref{puissance}, there exists $c\in\mathbf N$ such that $q_{cn}\ge q_n^4$ for every $n\in\mathbf N$. 
Let $\mathrm P:\mathrm L^2(\mathbf T,\mathbf R)\rightarrow\mathrm L^2(\mathbf T,\mathbf R)$ be the projector defined by 
$\mathrm Pu(x)=\sum_{k\le q_{ct}}\hat u(k)e^{2i\pi kx}$. Let $\mathrm Q = \mathrm{Id}-\mathrm P$.
One has $||\mathrm Qg||_{\mathrm L^2}\le \mathrm C/q_t$ and   $||\mathrm Qh||_{\mathrm L^2}\le \mathrm C/q_t$ ; 
indeed, by hypothesis 1), one has for example 
\begin{equation}\label{proveke}
||\mathrm Q h||_{\mathrm L^2}^2\le \frac{\mathrm{Var}^2(h)}{4\pi^2}\sum_{k:|k|\ge q_{ct}}\frac{1}{k^2}\le \mathrm C\frac{q_t^2}{q_{ct}}\le \frac{\mathrm C}{q_t^2}.
\end{equation}
But $|\int_0 ^1\phi g h \mathrm d x|\le |\int_0^1\phi\mathrm Pg\mathrm Ph\mathrm d x| + \int_0^1|\phi\mathrm Qg\mathrm Ph|\mathrm d x + \int_0^1|\phi g\mathrm Qh|\mathrm d x$. 
Because $\int_0^1\phi\mathrm d x= 0$, $||\phi ||_{\mathrm L^\infty }\le \mathrm{Var}(\phi)$. By hypothesis 1) and (\ref{proveke}), one has
\[
\int_0^1 |\phi g\mathrm Qh|\mathrm d x \le ||\phi ||_{\mathrm L^\infty }||g||_{\mathrm L^\infty }||\mathrm Qh ||_{\mathrm L^2 } \le \mathrm C .\mathrm{Var}(\phi)\frac{1}{q_t},
\]
and, because $||\mathrm Ph ||_{\mathrm L^2 }\le ||h||_{\mathrm L^2 }\le ||h||_{\mathrm L^\infty }$, one has similarly
\[
\int_0^1|\phi\mathrm Qg\mathrm Ph|\mathrm d x \le ||\phi ||_{\mathrm L^\infty }||\mathrm Q g ||_{\mathrm L^2 }||\mathrm P h ||_{\mathrm L^2 } \le \mathrm C . \mathrm{Var}(\phi)\frac{1}{q_t}. 
\]
Therefore, it will suffice to estimate $|\int_0^1\phi\mathrm Pg\mathrm Ph\mathrm d x|$ instead of $|\int_0 ^1\phi g h \mathrm d x|$. 

Next, let us prove that, if $t \ge c's$ for some $c'\in\mathbf N$, then the first inequality (\ref{blla}) implies the second one
(the proof of the first inequality makes obviously no use of this fact).
If $u,v\in\mathrm{BV}(\mathbf T,\mathbf R)$, one has $\mathrm{Var}(uv)\le ||u||_{\mathrm L^\infty }\mathrm{Var}(v) + ||v ||_{\mathrm L^\infty }\mathrm{Var}(u)$.
Therefore, using again the fact that $||\phi ||_{\mathrm L^\infty }\le \mathrm{Var}(\phi)$, hypothesis 1), and the first inequality (\ref{blla}), one gets
\[
|\int_0^1 (\phi g). h\mathrm d x|\le\mathrm C (||g||_{\mathrm L^\infty}\mathrm{Var}(\phi)+\mathrm{Var}(\phi)\mathrm{Var}(h))\frac{t^\beta}{q_t}
\le\mathrm C\frac{\mathrm{Var}(\phi)}{q_s}(\frac{q_st^\beta}{q_t}+\frac{q_s^2t^\beta}{q_t}).
\]
By Lemma \ref{puissance}, there exists $c'\in\mathbf N$ such that $q_{c'n}\ge q_n^2$ for every $n\in\mathbf N$. Therefore, if $t\ge c's$, one may write $t = c's + u$ with $u\ge 0$, and so, using (\ref{expcrois}),
\[
\frac{q_st^\beta}{q_t}+\frac{q_s^2t^\beta}{q_t} \le 
\frac{2 q_{c's}(c's+u)^\beta}{q_{c's+u}}\le \mathrm C\frac{(c's+u)^\beta}{e^{2\rho u}}\le \mathrm C c'^\beta s^\beta \frac{(1+u/cs)^\beta}{e^{2\rho u}} \le \mathrm Cs^\beta.
\]

One now comes to the proof itself (and one supposes $t < c's$). After some algebra one gets
\begin{eqnarray}
|\int_0^1 \phi\mathrm P g\mathrm P h\mathrm dx| &=& 
|\sum_{j,k : |j|,|k|\le q_{ct}}\hat g(j)\hat h(-k)\hat\phi(k-j)| \nonumber\\
&\le &
\frac{4\mathrm{Var}(\phi)}{2\pi}\sum_{1\le j,k\le q_{ct}, k\ne j}|\hat g(j)|.|\hat h(k)|.\frac{1}{|k-j|} + 
\frac{\mathrm{Var}(\phi)}{2\pi}\sum_{1\le k\le q_{ct}}\frac{|\hat g(k)|.|\hat h(k)|}{k}\nonumber\\
&\stackrel{\Delta}{=}& \frac{\mathrm{Var}(\phi)}{2\pi}(4\mathrm S_1 + \mathrm S_2). \label{suuum}
\end{eqnarray}

Let us estimate $\mathrm S_1$ given by (\ref{suuum}). 
The set $\Gamma_n$ ($n\ge 1$) are defined in (\ref{gamman}).
By (\ref{latticeinequality}) and (\ref{expcrois}), there exists a constant $w\ge 0$ such that, if $j\in\Gamma_{m+w}$, then $j>q_m$ ($m\ge 1$), and therefore
\begin{equation}\label{bankabou}
\mathrm S_1 = \sum_{1\le m,n\le ct+w}\sum_{\substack{j\in\Gamma_m,j\le q_{ct}\\k\in\Gamma_n,k\le q_{ct},k\ne j}}
|\hat g(j)|.|\hat h(k)|.\frac{1}{|k-j|} 
\stackrel{\Delta}{=}
\sum_{1\le m,n\le ct+w}\mathrm S (m,n).
\end{equation}
Let us fix $m,n\in\lbrace 1,\dots ,ct+w\rbrace$ and estimate $\mathrm S(m,n)$. 
Let us first consider the case $m\le n$. 
By (\ref{froulouk}), (\ref{Fourier}), (\ref{linearequ}), (\ref{ead})  and (\ref{gamman}), one has
\begin{equation}\label{lastig1}
|\hat g(j) |
= |\widehat{f_{ls}}(j) |\le \frac{|1 - e^{2i\pi b_s(l)q_sj\alpha }|}{|1 - e^{2i\pi j\alpha }|}
\frac{\mathrm{Var}(\psi) }{2\pi j }
\le
\frac{dj.|1-e^{2i\pi q_s\alpha }|}{1/q_m}\frac{\mathrm{Var}(\psi) }{2\pi j }
\le \mathrm C \frac{q_m}{q_s}.
\end{equation}
By (\ref{froulouk}), (\ref{Fourier}), the fact that $|1-e^{2i\pi x }|\le 2$ for all $x\in\mathbf R$, and (\ref{gamman}), one has
\begin{equation}\label{lastig2}
|\hat h(k) | 
= |\widehat{f_{lt}}(k) |
\le \frac{|1 - e^{2i\pi b_t(l)q_t k\alpha }|}{|1 - e^{2i\pi k\alpha }|}\frac{\mathrm{Var}(\psi) }{2\pi k }\le \mathrm C\frac{q_n}{k}.
\end{equation}
Therefore, by (\ref{bankabou}),  (\ref{lastig1}) and (\ref{lastig2}), and then (\ref{latticeinequality}), one has
\begin{eqnarray*}
\mathrm S(m,n) &\le& \mathrm C\frac{q_mq_n}{q_s}\sum_{\substack{j\in\Gamma_m,j\leq q_{ct}\\k\in\Gamma_n,k\le q_{ct},k\ne j}}\frac{1}{k|k-j|}
\le \mathrm C\frac{q_mq_n}{q_s}\sum_{k\in\Gamma_n,k\le q_{ct}}\frac{1}{k}\sum_{j\in\Gamma_m,j\le q_{ct},j\ne k}\frac{1}{|k-j|} \\
&\le&
\mathrm C\frac{q_mq_n}{q_s} \Big(
\frac{\mathrm C }{q_n}\sum_{1\le u \leq q_{ct} }\frac{1}{u}
\Big) .\Big(
\frac{\mathrm C }{q_m}\sum_{1\le u \le q_{ct} }\frac{1}{u}
\Big)
\le
\frac{\mathrm C}{q_s}\ln^2q_{ct}.
\end{eqnarray*}
The case $m\ge n$ is analogous : one uses the estimates $|\hat g(j)|\le\mathrm Cq_m/j$ and $|\hat h(k)|\le\mathrm C q_n/q_t$, to obtain $\mathrm S(m,n)\le (\mathrm C/q_t)\ln^2q_{ct}$. 
Therefore, one has $\mathrm S_1\le \mathrm C(ct+w)^2(\ln^2q_{ct})/q_s$.

The sum $\mathrm S_2$ is estimated in the same way. One gets $\mathrm S_2\le \mathrm C (ct + w)(\ln q_{ct})/q_s$.
To get the result, one uses then the inequality $q_{ct}\le (2d)^{ct}$, where $d$ is given by (\ref{ead}), and one takes $\beta = 4$. $\square$

Let us show that hypothesis 3) is satisfied. 
\begin{Lemma}\label{majore}
Let $\alpha$ be a number of constant type.
Let $n\ge 0$ and $0\le m\le q_n$. 
One has $||y_m||_{\mathrm L^2}\le\mathrm C\sqrt n$.
\end{Lemma}
\emph{Proof.} 
By Lemma \ref{puissance}, there exists $c\in\mathbf N$ such that $q_{cn}\ge q_n^2$ for every $n\in\mathbf N$.
By (\ref{Fourier}), one has
\begin{equation}\label{kuru}
\int_0^1 y_m^2\mathrm d x \le
\frac{\mathrm{Var}^2(\psi)}{2\pi^2}\sum_{k=1}^{q_{cn}-1}\frac{1}{k^2}\frac{|1-e^{2i\pi mk\alpha}|^2}{|1-e^{2i\pi k\alpha}|^2}
+\frac{1}{2\pi^2}\sum_{k\ge q_{cn}}\frac{\mathrm{Var}^2(y_m)}{k^2}.
\end{equation}
Because $y_m = \sum_{j=0}^{m-1}\mathrm R_{j\alpha }\psi$ ($m\ge 1$), one has
$\mathrm{Var}(y_m)\le \mathrm{Var}(\psi)m\le \mathrm{Var}(\psi)q_n$, 
and thus the second term in (\ref{kuru}) is bounded by a constant.  
The sets $\Gamma_m$ $(m\ge 1)$ are defined in (\ref{gamman}). 
By (\ref{latticeinequality}) and (\ref{expcrois}), there exists a constant $w\ge 0$ such that, if $j\in\Gamma_{m+w}$, then $j>q_m$ ($m\ge 1$), and so
\[
\sum_{k=1}^{q_{cn}-1}\frac{1}{k^2}\frac{|1-e^{2i\pi mk\alpha}|^2}{|1-e^{2i\pi k\alpha}|^2} \le
\sum_{m=1}^{cn+w}\sum_{k\in\Gamma_m}\frac{1}{k^2}\frac{2}{|1-e^{2i\pi k\alpha}|^2} \le
\sum_{m=1}^{cn+w}\sum_{k\in\Gamma_m}\frac{2q_m^2}{k^2}\le \mathrm C \sum_{m=1}^{cn+w}\sum_{u\ge 1 }\frac{2q_m^2}{u^2q_m^2}
\le \mathrm Cn. \quad\square
\]
One has $\mathrm S_n = z_{s_n} = y_m$ for some $m\le  q_{n+1}$, and therefore, by Lemma \ref{majore}, $||\mathrm S_n||_{\mathrm L^2 }\le \mathrm C\sqrt{n+1}\le \mathrm C\sqrt n$.
The estimate $||\mathrm S_n||_{\mathrm L^2}\ge \epsilon\sqrt n$ is obtain by the next Lemma.
\begin{Lemma}\label{minore}
Under the hypotheses of Proposition \ref{result3}, there exits $\epsilon >0$ such that $||z_{s_n}||_{\mathrm L^2 }\ge \epsilon \sqrt n$ for every $n\ge 1$.
\end{Lemma}
\emph{Proof.}
Let $\delta >0$. 
By (\ref{expcrois}), there exists $l\ge 1$ such that, for each $n\ge 0$, $q_n/q_{n+l}\le \delta$.
Therefore, for each $k\ge 0$, $q_n/q_{n+kl}\le \delta^k$.
Now let us construct a sequence $(t_n)_{n\ge1 }\subset\mathbf N$.
For $n\ge 1$, $t_n = 0$ except in the following cases.  
If $n=4kl$ for some $k\ge 1$, then, by point 4) of Lemma \ref{fraccont},
there exists $m\in\lbrace n,\dots ,n+3 \rbrace$ such that $q_m$ is odd and that $q_m|q_m\alpha |_\mathbf T < 1/2$ ; 
one sets then $t_m = q_m$ (and one takes the smallest $m$ if there is more than one possibility).

Let us fix $n\ge 1$.
Because, $q_0 + \dots  +q_n\le q_{n+3}$, one has $t_1 + \dots + t_n \le q_{n+3}$.
Therefore, if $r = t_1 +\dots + t_n$, one has $||z_{s_{n+3}}||_{\mathrm L^2 }\ge ||z_{q_{n+3}}||_{\mathrm L^2 }\ge ||y_{r}||_{\mathrm L^2 }$. 
But, by (\ref{Fourier}), (\ref{psistar}) and (\ref{fundamentalFracCont}), one has  
\begin{eqnarray*}
||y_{r}||_{\mathrm L^2 }^2 &=& 
\frac{8}{\pi^2}\sum_{k\ge 1, k \mathrm{odd}}\frac{1}{k^2}\frac{|1-e^{2i\pi r k\alpha}|^2}{|1-e^{2i\pi k\alpha}|^2}
\ge \frac{8}{\pi^2 }\sum_{1\le u\le n : q_u\mathrm{odd}}\frac{1}{q_u^2}\frac{|1-e^{2i\pi r q_u \alpha}|^2}{|1-e^{2i\pi q_u \alpha}|^2} \\
&\ge&
\mathrm C \sum_{1\le u\le n:q_u\mathrm{odd} } |1- e^{2i\pi r q_u\alpha }|^2 
\ge \mathrm C \sum_{u=1}^n |rt_u\alpha |^2_{\mathbf T }.
\end{eqnarray*}

Let $u\in\lbrace 1,\dots ,n \rbrace$ be such that $t_u\ne 0$ (and thus $t_u =q_u$).
One writes $rt_u\alpha = t_u^2\alpha + \tau$, where 
$\tau = (t_1 + \dots + t_{u-1})t_u\alpha + t_u(t_{u+1}\alpha + \dots + t_n\alpha)$.
By (\ref{triangle}), one has $|\tau |_\mathbf T\le (t_1 + \dots +t_{u-1})|t_u\alpha |_\mathbf T + t_u (|t_{u+1}\alpha |_\mathbf T + \dots + |t_n\alpha |_\mathbf T)$.
Let us now adopt the convention that $1/t_n=0$ when $t_n=0$. One has
\begin{equation}\label{provprov}
|\tau |_{\mathbf T } \le \frac{t_1}{q_u} + \dots + \frac{t_{u-1}}{q_u} + \frac{q_u}{t_{u+1}} + \dots + \frac{q_u}{t_n} \le 2\sum_{k=1}^\infty \delta^k.
\end{equation}
So, taking $\delta$ small enough (and thus $l$ big enough), 
$|\tau |_\mathbf T$ can be made arbitrarily small.

If $p\in\mathbf N$, if $x\in\mathbf R$ and if $p|x|_{\mathbf T}\le1/2$, then $|px|_\mathbf T=p|x|_\mathbf T$. 
Therefore, if $d$ is given by (\ref{ead}), one has by (\ref{fundamentalFracCont}), and because $t_u|t_u\alpha |_{\mathbf T }<1/2$ by construction, that
\[ 
|t_u^2\alpha |_\mathbf T = q_u|q_u\alpha |_\mathbf T \ge \frac{q_u}{q_u + q_{u+1}}\ge \frac{1}{3d}.
\]
Therefore, there exists $\mathrm C>0$ such that $|rt_u\alpha |^2_\mathbf T \ge \mathrm C$, 
and so $||y_r||^2_{\mathrm L^2}\ge \mathrm C n/l$. $\square$ 

This ends the proof of Proposition \ref{result3}. $\square$

\noindent\emph{Proof of Corollary \ref{result2}.}
For $j$ large enough, one has $n_j\ge 1$, and 
one may thus find $\tau (j)\in\mathbf N$ such that $q_{\tau(j)} \le n_j < q_{\tau(j)+1 } $.
Now, by Lemma \ref{majore} and the hypothesis on $(\sigma_j)_{j\ge 1 }$, one has
$||y_{n_j}/\sigma_j\sqrt j ||_{\mathbf L^2 }\le \mathrm C \sqrt{(\tau(j)+1 )/j}$.
Because 
$y_{n_j}/\sigma_j\sqrt j\stackrel{\mathrm D}{\rightarrow }\mathrm G(1)$, 
there has to be a number $\mathrm C>0$ such that $\tau (j)\ge \mathrm C j$ for $j$ large enough.
The result follows from the fact that $n_j\ge q_{\tau(j) }$ and that $(q_n)_{n\ge 0 }$ grows exponentially with $n$. $\square$

\noindent\emph{Acknowledgments.} I am very grateful to Professors J. Bricmont and A. Kupiainen for useful discussions and comments.  
I thank Belgian Interuniversity Attraction Poles Program for financial support.

\end{document}